\documentclass[journal,10pt,letterpaper]{IEEEtran}
\usepackage{amsmath}
\usepackage{amsthm}
\usepackage{amsfonts}
\usepackage{amssymb}
\usepackage{makeidx}
\usepackage{cite}
\usepackage{graphicx,bigints}
\usepackage{mathtools}
\usepackage{cases}
\usepackage{color}
\usepackage{hyperref}
\usepackage{bbm}
\usepackage[normalem]{ulem}
\usepackage{braket}
\usepackage{textcomp,gensymb}

\usepackage{enumitem}

\usepackage{epstopdf}
\usepackage{xcolor}
\usepackage{lipsum}

\usepackage{multirow}

\makeatletter
\DeclareFontFamily{U}{tipa}{}
\DeclareFontShape{U}{tipa}{m}{n}{<->tipa10}{}
\newcommand{\arc@char}{{\usefont{U}{tipa}{m}{n}\symbol{62}}}%

\newcommand{\arc}[1]{\mathpalette\arc@arc{#1}}

\newcommand{\arc@arc}[2]{%
	\sbox0{$\m@th#1#2$}%
	\vbox{
		\hbox{\resizebox{\wd0}{\height}{\arc@char}}
		\nointerlineskip
		\box0
	}%
}
\makeatother

\newtheorem{corollary}{Corollary}
\newtheorem{lemma}{Lemma}

\newtheorem{definition}{Definition}

\DeclareMathOperator{\cC}{\mathcal{C}}

\DeclareMathOperator{\cO}{\mathcal{O}}

\DeclareMathOperator{\cL}{\mathcal{L}}
\DeclareMathOperator{\cS}{\mathcal{S}}

\DeclareMathOperator{\bR}{\mathbb{R}}

\DeclareMathOperator{\bP}{\mathbf{P}}

\DeclareMathOperator{\bE}{\mathbf{E}}

\newcommand*\diff{\mathop{}\!\mathrm{d}}

\newcommand*\nnb{\nonumber}

\newcommand{\overbar}[1]{\mkern 1.5mu\overline{\mkern-1.5mu#1\mkern-1.5mu}\mkern 1.5mu}

\newcommand\independent{\protect\mathpalette{\protect\independenT}{\perp}}
\def\independenT#1#2{\mathrel{\rlap{$#1#2$}\mkern2mu{#1#2}}}

\newcommand{\ea}{\stackrel{(\text{a})}{=}}
\newcommand{\eb}{\stackrel{(\text{b})}{=}}

\definecolor{sandy}{HTML}{E6E2AF}
\definecolor{stone}{HTML}{A7A37E}
\definecolor{beach}{HTML}{EFECCA}
\definecolor{ocean}{HTML}{046380}
\definecolor{diver}{HTML}{002F2F}

\definecolor{Firenze1}{HTML}{468966}
\definecolor{Firenze2}{HTML}{FFF0A5}
\definecolor{Firenze3}{HTML}{FFB03B}
\definecolor{Firenze4}{HTML}{B64926}
\definecolor{Firenze5}{HTML}{8E2800}
\definecolor{mediumpersianblue}{rgb}{0.0, 0.4, 0.65}
\definecolor{hongik}{HTML}{004498}
\definecolor{cobalt}{rgb}{0.0, 0.28, 0.67}
\definecolor{burntorange}{rgb}{0.8, 0.33, 0.0}

\definecolor{ultramarineblue}{rgb}{0.25, 0.4, 0.96}

\title{Cox Point Processes for Multi Altitude LEO Satellite Networks}

\author{Chang-Sik Choi,~\IEEEmembership{Member,~IEEE,}~and~Fran\c{c}ois~Baccelli,~\IEEEmembership{Member,~IEEE}
		\IEEEcompsocitemizethanks{\IEEEcompsocthanksitem{Chang-Sik Choi is with Dept. of EE, Hongik University, South Korea. François Baccelli is with Inria Paris and Telecom Paris, France.  (email: chang-sik.choi@hongik.ac.kr, francois.baccelli@inria.fr)}}
\IEEEcompsocitemizethanks{\IEEEcompsocthanksitem{Last revised: \today}}
}
\begin{document}
	
	\maketitle 
	
	\begin{abstract}
To model existing or future low Earth orbit (LEO) satellite networks leveraging multiple constellations, we propose a simple analytical approach to represent the clustering of satellites on orbits. More precisely, we develop a variable-altitude Poisson orbit process that effectively captures the geometric fact that satellites are always positioned on orbits, and these orbits may vary in altitude. Conditionally on the orbit process, satellites situated on these orbits are modeled as linear Poisson point processes, thereby forming a Cox point process. For this model, we derive useful statistics, including the distribution of the distance from the typical user to its nearest visible satellite, the outage probability, the Laplace functional of the proposed Cox satellite point process, and the Laplace transform of the interference power from the Cox-distributed satellites under general fading. The derived statistics enable the evaluation of the performance of such LEO satellite communication systems as functions of network parameters.
	\end{abstract}

	\section{Introduction}
	\subsection{Motivation and Background}
	\IEEEPARstart{L}{EO} satellites provide global connectivity to millions of devices on Earth \cite{8700141,8626457,38821}. The applications of LEO satellite networks are numerous \cite{8700141}: they provide Internet connections to  devices where ground infrastructure is unavailable; they enable localization and emergency communications for aerial and ground devices; they also provide cheaper Internet connections to developing countries. They can even be integrated with terrestrial networks to enable reliable connections to devices \cite{38821}. To support these applications, LEO satellite networks must have a very large number of satellites.
	
	\par The viability and performance of LEO satellite communications are determined by the way satellites are distributed in space. Various evaluation methodologies have been proposed to obtain the performance of such networks. For satellite layout, some studies used probabilistic approaches including the binomial point process approach \cite{9079921,9218989,9497773,9678973}. In contrast to the simulation-based approach, the benefits of employing such analytical models lie in the fact that they allow one to predict large-scale network behavior in functions of a few key parameters such as the mean number of satellites, their altitudes, etc. Nevertheless, binomial satellite point processes are not able to incorporate the fact that the satellites are located on approximately circular trajectories around the Earth, namely their orbits. In this paper, we provide a tractable model that incorporates this fact in the multi-altitude LEO satellite case, by generalizing the work in \cite{choi2022analytical} where all orbits are at the same altitude. Specifically, we present an analytical framework leveraging a Cox point process so that orbits are first created according to a Poisson point process on a cuboid and then satellites are distributed as Poisson point processes conditionally on these orbits. We derive key statistical properties of the proposed network model that are critical to derive the performance of such satellite networks as functions of the altitude distribution, the mean number of orbits, the mean number of satellites per orbit, and the distribution of fading. 
	

	
	\subsection{Contributions}
	\textbf{Modeling of multi altitude LEO satellite constellations}:  By developing a nonhomogeneous Poisson point process  in a cuboid, we create an isotropic Poisson orbit process of orbits in the Euclidean space. Then, conditionally on the orbit process, satellites are distributed as linear Poisson point processes of mean $ \mu $ on these orbits. Our motivation is to represent general LEO satellite networks where satellites are located at different altitude bands. The developed Cox satellite model is proved to be isotropic, namely invariant by all rotations of the reference space. 
	
			{Our study aims to understand essential performance-related statistics of the LEO satellite networks consist of multiple constellations characterized by different geometric parameters including their altitudes. In practice, a given constellation plan determines the number of orbital planes, the inclinations of orbits, their altitudes, and the number of satellites per plane. However, the actual geometry of each LEO satellite constellation may deviate from this plan significantly due to practical reasons like satellite retirement, new or delayed satellite launches, and satellite decay. In addition, the reduction in launch costs has facilitated the emergence of numerous LEO satellite constellations, and this trend is expected to continue, making the structure of the collection of all constellations increasingly complex. As a result, obtaining precise information about all satellites' locations and inclinations becomes highly challenging, which justifies considering the random modeling approach considered in this paper.} 	
		
	\textbf{Statistical properties of the proposed Cox point process}: Isotropy ensures that the statistical properties of the network seen from all points are the same. Leveraging this feature, we derive the probability distribution function for the distance between the typical user and its nearest visible satellite and subsequently derive the outage probability of the proposed network model. Using these results, we derive the Laplace functional of the proposed satellite Cox point process and from it, we evaluate an integral expression for the Laplace transform of total interference power for a general fading scenario. These formulas are directly employed to assess network performance metrics, such as the Signal-to-Interference-plus-Noise Ratio (SINR) of the typical user in future LEO satellite networks consisting of multiple constellations, each characterized by distinct geometric parameters. Finally, as an example illustrating the application of the proposed model, we compare the coverage probability of the proposed Cox model with that of the future Starlink 2A constellation and an existing binomial-based LEO satellite model. This comparison reveals that the proposed Cox model has the potential to better represent existing or future LEO satellite constellations.

	\section{System Model}

	\subsection{Cox Satellite Point Process}

The center of the Earth is $ O=(0,0,0) $. The Earth radius is denoted by $ r_e. $ The $ xy $-plane is the equatorial plane and the $ x $-axis is the longitude reference direction. In this paper, we only focus on a snapshot of the network geometry.

	 Consider a cuboid $ \cC=[r_a,r_b]\times[0,\pi)\times [0,\pi) $ where $ r_a, r_b$ are the minimum and maximum altitudes, respectively. Consider a Poisson point process $ \Xi $ of intensity measure $ \frac{\lambda\sin(\phi)}{2\pi}\nu(\diff \rho) \diff \phi \diff \theta$ on the cuboid $ \cC $. We assume $ \int_{r_a}^{r_b}\nu(\diff\rho)=1. $ Then, we build an orbit process by mapping each point of $ \Xi $, say $ {(\rho,\theta,\phi)}$ into an orbit $ l(\rho,\theta,\phi) $ in the Euclidean space. Specifically, the first coordinate $ \rho $ gives the orbit's radius, whereas $ \theta $ and $\phi$ give the orbit's longitude and inclination, respectively. For the Poisson point process on the cuboid, we write $ \Xi = \sum_{i} \delta_{Z_i}, $ where $ Z_i $ is the location of the point of $ \Xi $. 
Since there are on average $ \lambda $ points of $ \Xi $, there are on average $ \lambda $ orbits. The orbit process $ \cO $ in $ \bR^3 $ is 
\begin{equation}
	\cO = \bigcup_{Z_i\in\Xi}{l(\rho_{i},\theta_{i},\phi_{i})} .
\end{equation}
		\par
\begin{figure}
	\centering
	\includegraphics[width=1\linewidth]{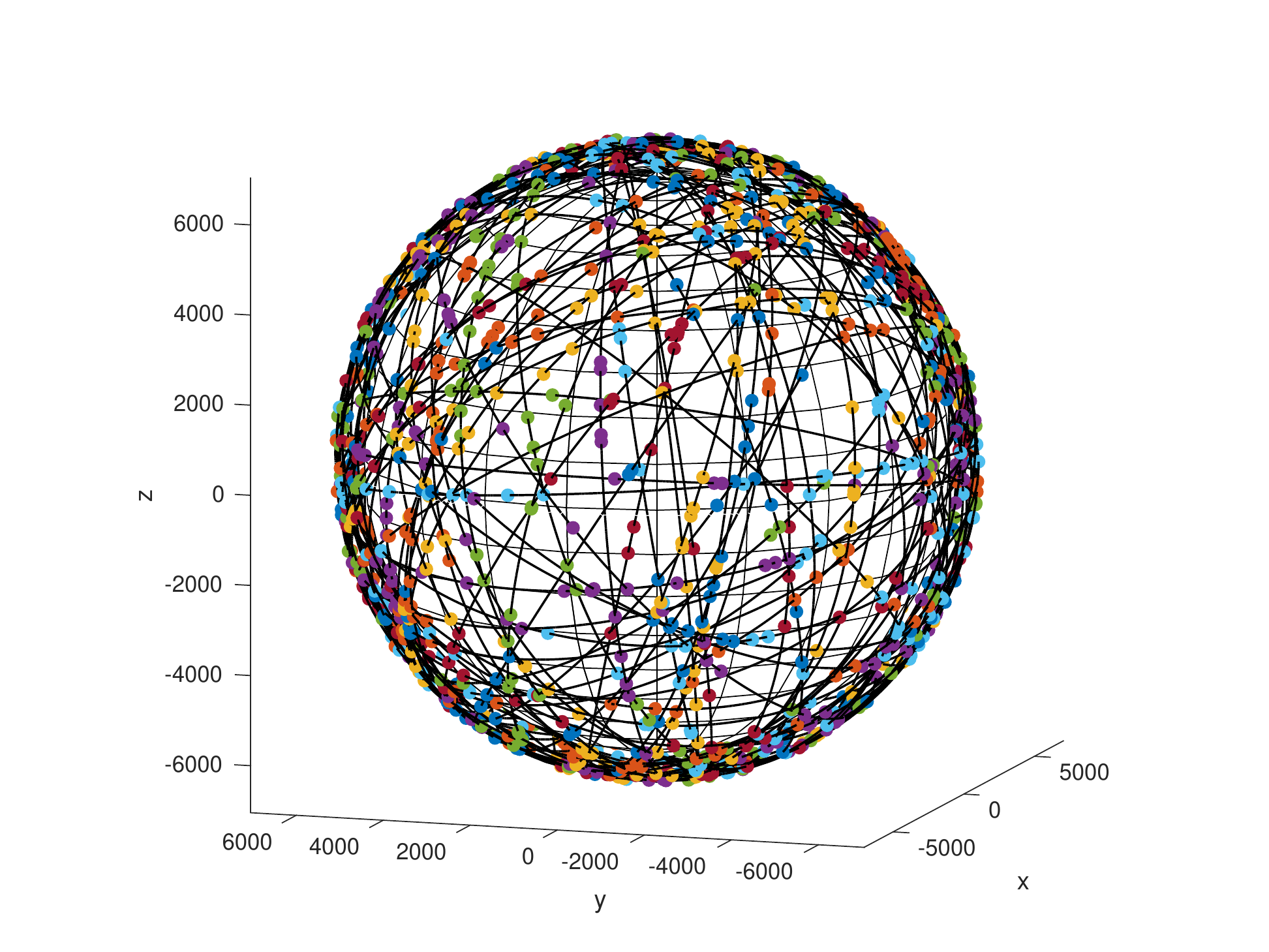}

	\caption{The proposed Cox satellite model with $ r_a=7000 $ km, $ r_b=7050 $ km. We use  $ \lambda=72$, $ \mu=22$, and $ \nu(\diff \rho) = {\diff\rho}/{(r_b-r_a)} $.}
	\label{fig:604070007100}
\end{figure}

\begin{figure}
	\centering
	\includegraphics[width=1\linewidth]{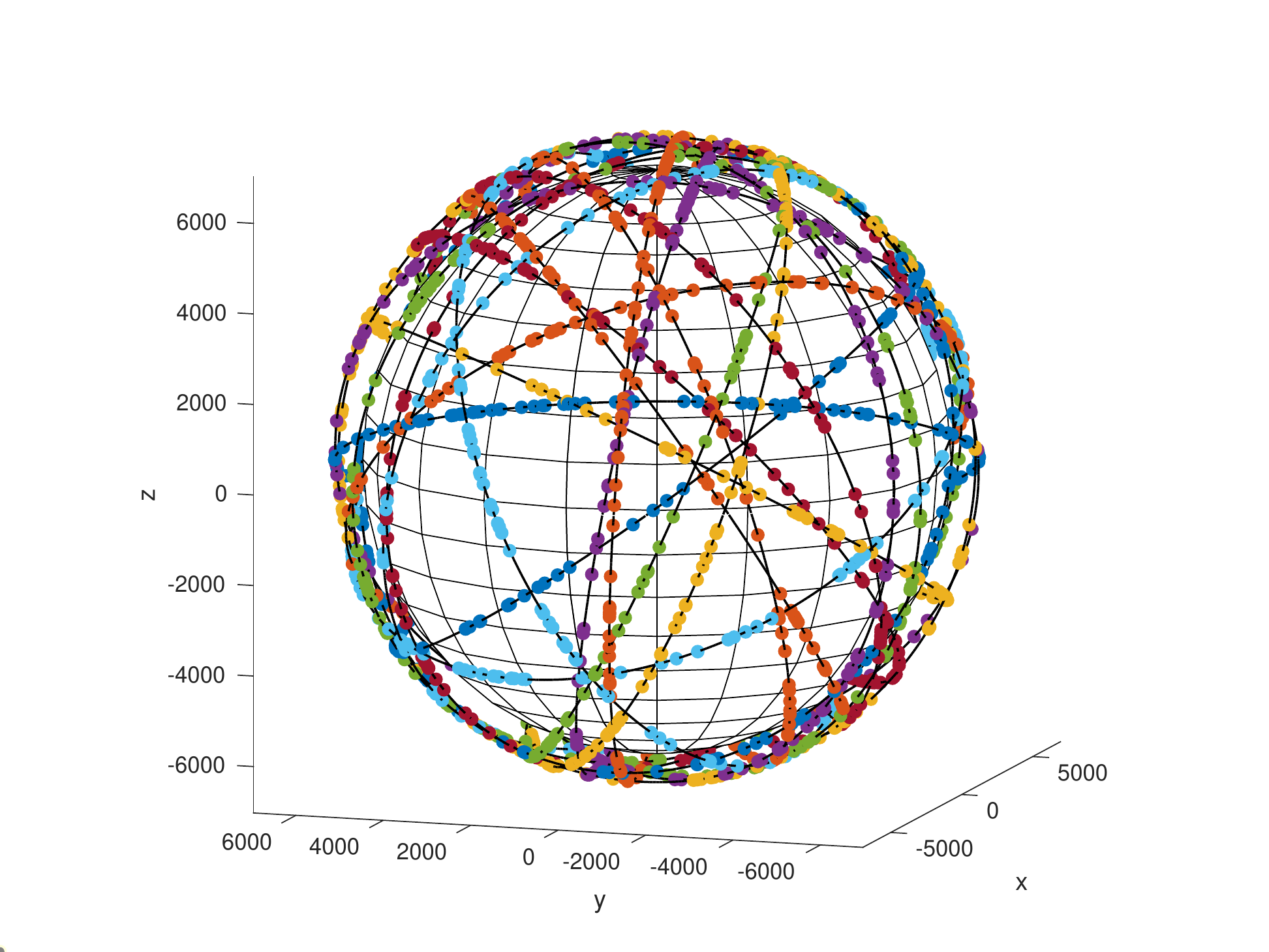}
	\caption{The Cox-modeled satellite with $ r_a=7000$ km and $r_b=7050$ km. We use $ \lambda=25 $, $ \mu=100 $, and $ \nu(\diff \rho) = {\diff\rho}/{(r_b-r_a)} $.}
	\label{fig:3060_70007500_3}
\end{figure}

 Conditionally on $ \Xi, $ the locations of satellites on each orbit $ l(\rho_{i},\theta_{i},\phi_{i}) $ are modeled as a homogeneous Poisson point process $ \psi_i $ of intensity $ \mu/(2\pi\rho_{i}) $ on this orbit. Equivalently, the orbital angles of satellites on each orbit are modeled as a 1-dim homogeneous Poisson point process $ \overbar{\psi}_{i} $ of intensity $ \mu/(2\pi)$ on $ [0,2\pi) $. Since the satellites are distributed conditionally on $ \Xi $, the satellite point process $ \Psi $ is a Cox point process denoted as 
		\begin{equation}
			\Psi = \sum_{i} \psi_{i}.
		\end{equation}
		Figs. \ref{fig:604070007100} --\ref{fig:3060_70007500_3} depict the proposed satellite Cox point process. In the figures, we use $ \nu(\diff\rho)=\frac{\diff\rho}{r_b-r_a} $, i.e., the radii of orbits are uniformly distributed in $ (r_a,r_b). $ The proposed model can be used to represent, for example, LEO satellite networks consisting of multiple operators with orbits at different altitudes.	The case of all satellites located at the same altitude in \cite{choi2022analytical} is a special case of the proposed model with $ \nu(\diff \rho) = \delta_{r_a}(\diff\rho) $ where $ r_a $ is the radius of orbits.

	\subsection{User Distribution}
	We assume that users are uniformly distributed on the surface of Earth $ \{(x,y,z)|x^2+y^2+z^2=r_e^2\} $ and their locations are independent of those of LEO satellites.

\section{Results}
	\subsection{Statistical Results}
	In this section, we derive/establish (i) the mean number of LEO satellites, (ii)  the isotropy of $ \Psi $, (iii) the distribution of the distances from the LEO satellites to an arbitrarily located user, (iv) the distribution of the distance to the nearest visible satellite, (v) the outage probability, (vi) the Laplace functional of $ \Psi $, and (vii) the Laplace transform of the total interference under general fading. These statistical properties directly determine the performance of downlink LEO satellite communications. 
			\begin{lemma}
			The average number of the proposed Cox satellite point process  is $ \lambda\mu. $
		\end{lemma}
		\begin{IEEEproof}
			Let $\cS$ be the set of all spheres with radii from $r_a$ to $r_b.$ The average number of satellites is given by 
			\begin{align*}
				\bE\left[\Psi(\cS)\right] 
				&= \bE\left[\left.\sum_{Z_i\in\Xi}\bE\left[\sum_{X_j\in\psi_{i}} 1\right|\Xi\right]\right]\\
				&=\bE\left[\left.\sum_{Z_i\in\Xi} \int_{0}^{2\pi}\frac{\mu}{2\pi} \diff x \right|\Xi\right]\\
				&=\mu\int_{0}^\pi \int_0^\pi \int_{r_a}^{r_b}\frac{\lambda\sin(\phi)}{2\pi}\nu(\diff \rho) \diff \theta \diff \phi= \lambda\mu, 
			\end{align*}
			where we used Campbell's mean value theorem \cite{baccelli2010stochastic}.
		\end{IEEEproof}
	
	Below we show that $ \cO $ is invariant w.r.t. rotations. This allows one to evaluate the performance of network seen by a typical user at the north pole. 
	\begin{lemma}\label{lemma:2}
		The distributions of $ \cO $ and $ \Psi $ are isotropic.
	\end{lemma}
\begin{IEEEproof}
Let $\mathbb{S}$ be the unit sphere of center $O$.  Let $ U $ be a uniformly distributed random point on $\mathbb{S}$.  For each ${U}$, there is a unique directed orbit $\mathbb{O} (U)\subset \mathbb{S}$ the orbit whose right-hand rule normal vector is $\overrightarrow{OU}$. Note that $\mathbb{O} (U)$ is a factor of ${U}$; namely, for all $\bR^3$ rotations $R$ of center $o$, the orbit $\mathbb{O} (R(U))$ coincides with the orbit $R(\mathbb{O} (U))$. Since the law of $U$ is isotropic on $\mathbb{S}$, it follows from the relation $\mathbb{O} (R(U))=R(\mathbb{O} (U))$ that the law of $\mathbb{O} (U) $ is also isotropic.

	It is well known that the uniform vector ${U}$ can be represented as ${U} = (\sqrt{1-V^2}\cos(\Theta), \sqrt{1-V^2}\sin(\Theta), V)
$
	with $V\sim\text{Uniform}(-1,1)$,  $\theta\sim\text{Uniform}(0,2\pi),$ and $V\independent \Theta.$  
	
	For the directed orbit $\mathbb{O} (U)$, let $\theta\in [0,2\pi)$ to be the angle of the ascending point from the $x$-axis and $\phi\in [0,\pi)$ to be the azimuth of the normal vector. Then, $\phi$ coincides with the inclination.  
Using the distributions of $V$ and $\Theta$, 
	\begin{align}
		\phi = \arccos(V)	 \ \text{ and } \theta = \Theta+\pi/2 \mod 2\pi.\label{3}
	\end{align}
	Since $V\independent \Theta,$ we get $\phi\independent \theta$. 
	Using Eq. \eqref{3},  we have $\theta \sim \text{Uniform}[0,2\pi)$
	and the PDF of $\phi$ is given by 
	\begin{equation}
		f_{\phi}(x) =
			\frac{\sin(x)}{2} \text{ for }0\leq x< \pi \text{ and }	0  \text{ otherwise}.
	\end{equation}
	The isotropic directed orbit Poisson point process can hence be represented as a Poisson point process of density $\Lambda( \theta, \phi) =\frac{ \lambda}{4\pi}\sin(\phi) $ on the rectangle set $\mathcal{R} = [0,\pi)\times [0,2\pi)$.
	Here, $\lambda$ is the mean number of directed orbits. 
	
	Furthermore, the directed orbit with angles $(\theta,\phi)$ and the directed orbit with angles $(\theta+\pi, 
	\phi+\pi/2 \mod \pi)$ reduce to the same orbit by forgetting the orbit direction.
	Therefore, for the undirected isotropic orbit, its longitude angle
$\widetilde \theta$ is uniformly distributed in $[0,\pi).$ The isotropic undirected orbit Poisson point process can hence be represented as a Poisson point process
	of density $\widetilde{\Lambda}(\widetilde \phi, \widetilde \theta) = \frac{ \lambda}{2\pi}\sin(\widetilde \phi) $ on the rectangle $\widetilde{\mathcal{R}} = [0,\pi)\times [0,\pi)$. 
	
	Finally, since the proposed orbit process intensity measure is defined as the product form $\frac{\lambda\sin(\phi)}{2\pi}\nu(\diff \rho)\diff \theta \diff \phi $, we have that $\cO$ is isotropic and so is $\Psi $.
\end{IEEEproof}

\begin{lemma}\label{lemma:3}
	Consider a satellite $ X $ of orbital angle $ \omega_j $ on the orbit $ l(\rho_{i}, \theta_{i},\phi_{i}) $. The distance from $ (0,0,r_e) $ to the satellite, denoted by $ X(\rho_{i},\theta_i,\phi_i,\omega_j) $, is 
	\begin{equation}
		\sqrt{\rho_i^2-2\rho_ir_e\sin(\omega_j)\sin(\phi_i)+r_e^2}.
	\end{equation}
\end{lemma} 
\begin{IEEEproof}
		The coordinates $ (x,y,z)\in\bR^3 $ of the satellite that has the orbital angle $ \omega_j $  on the orbit $ l(\rho_{i},\theta_{i},\phi_{i}) $ are given by 
	\begin{align}
		x &= \!\sqrt{\rho_i^2\cos^2(\omega_j)+ \rho_i^2\sin^2(\omega_j)\cos^2(\phi_{i})}\cos\left(\tilde{\theta}+\theta_{i}\right),\label{x}\\
		y &= \!\sqrt{\rho_i^2\cos^2(\omega_j)+ \rho_i^2\sin^2(\omega_j)\cos^2(\phi_{i})}\sin\left(\tilde{\theta}+\theta_{i}\right),\label{y} \\
		z &= \rho_i\sin(\omega_j)\sin({\phi}_{i}),\label{z}\\
		\tilde{\theta} &= \arctan\left({\tan(\omega_j)\cos({\phi_{i}})}\right).
	\end{align}
	As a result, the distance from $ (0,0,r_e) $ to the satellite is 
	\begin{align*}
		\|(x,y,z)-(0,0,r_e)\| 
		&= \sqrt{\rho_i^2-2\rho_ir_e\sin(\omega_j)\sin(\phi_i)+r_e^2}.
	\end{align*}
	Note that this distance is independent of the variable $ \theta. $ 
\end{IEEEproof}		

	 Since (i) users are independent of $ \Psi $ and (ii) $ \Psi $ is invariant by rotations (Lemma \ref{lemma:2} ), one can consider a typical user at $ (0,0,r_e) $ and study the distribution of the network performance it experiences, which represents that of all users in the network.
	 
\begin{figure}
	\centering
	\includegraphics[width=0.7\linewidth]{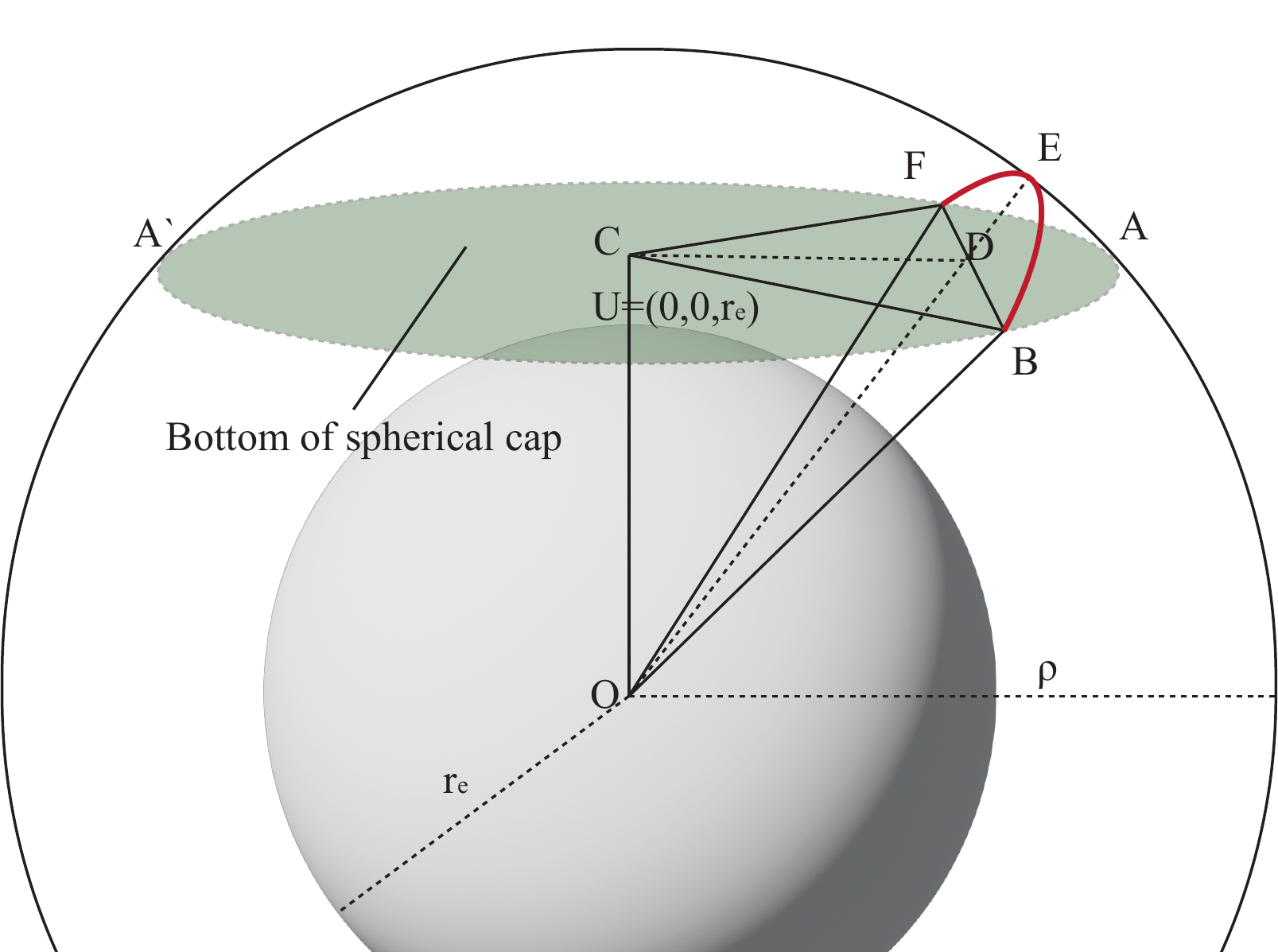}
	\caption{The arc of orbit $ l(\rho,\theta,\phi) $ in spherical cap $ C(\rho,d) $.}
	\label{fig:shortarclength}
\end{figure}

Let $ C(d) $ be set of all points on $ \cS $ whose distances from the typical observer $ u $ are less than a distance $ d $ and which are visible from $(0,0,r_e).$ For any $ r_a\leq \rho \leq r_b, $ we have 
\begin{align}
	C(d) &= \bigcup_{r_a\leq \rho\leq r_b}C(\rho, d)\nnb,
\end{align}
where the spherical cap associated with radius $\rho$ is given by 
\begin{align} 
	C(\rho, d)&=\cup_{r_a\leq \rho\leq r_b}\left\{(x,y,z)\in\bR^3|z\geq r_e,\right.\nnb\\
	&x^2+y^2+z^2=\rho^2,\left.{x^2+y^2+(z-r_e)^2}\leq d^2\right\}.\nnb
\end{align}
 See Fig. \ref{fig:shortarclength} for $C(\rho, d)$.

%
\begin{lemma}
	The length of the arc given by the intersection of the spherical cap $ C (\rho,d) $ and the orbit $ l(\rho,\theta,\phi) $ is 
	\begin{equation}
		2\rho\arcsin\left(\sqrt{1-\left(\frac{\rho^2+r_e^2-d^2}{2\rho r_e\sin(\phi)}\right)^2}\right),
	\end{equation}
for $ \rho - r_e \leq d\leq \sqrt{\rho^2-r_e^2} $. 
\end{lemma}

\begin{figure}
	\centering
	\includegraphics[width=1\linewidth]{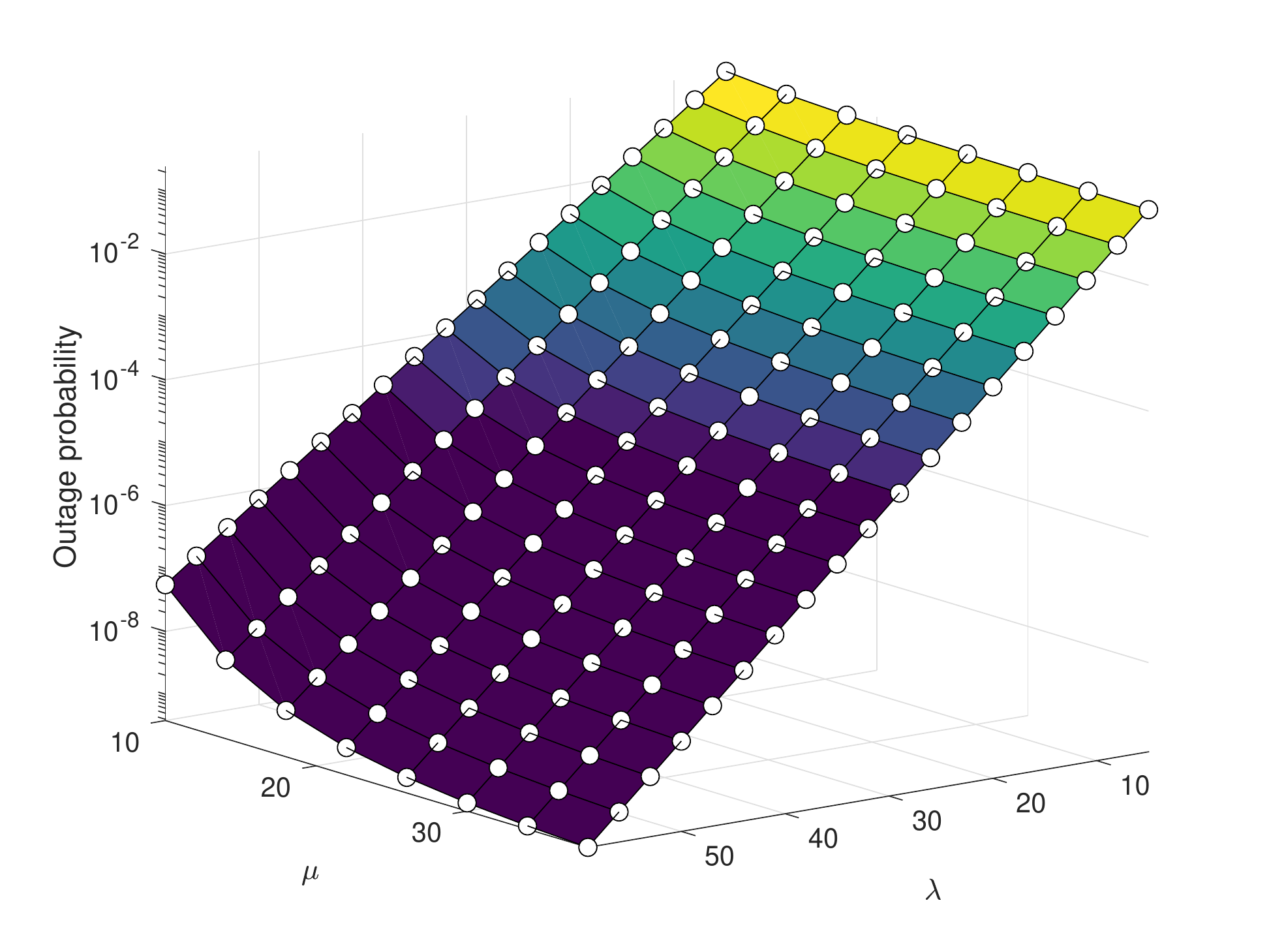}
	\caption{The outage probability with $ r_a=7000 $ km and $ r_b=8500 $ km. We use $ \lambda=72, $ $ \mu=22,  $ and $ \nu(\diff \rho) = {\diff\rho}/{(r_b-r_a)} $.}
	\label{fig:outagevariable70007500}
\end{figure}

\begin{IEEEproof}
Consider $ C(\rho,d). $ Let $ \xi $ be the angle $ \angle AOU$  in Fig. \ref{fig:shortarclength}. Then, we use the law of Cosine to obtain $\cos(\xi) = {(\rho^2+r_e^2-d^2)}/{(2\rho r_e)}. $

\par For the triangle $ \triangle BCD,$  we have $  \overbar{CD}= \rho\cos(\xi)\cot(\phi)$. 
Since the angle $ \angle BDC$ is $ \pi/2$, we obtain 
\[\overbar{BD} = \sqrt{\rho^2\sin^2(\xi) - \rho^2 \cos^2(\xi)\cot^2(\phi)}. \] 
For $ \triangle BOD $, $ \overbar{OB} = \rho $ and let $ {\kappa}' = \angle BOD $. Then we have 
\begin{equation*}
	\sin(\kappa')={\overbar{BD}}/{\rho}=  \sqrt{\sin^2(\xi) - \cos^2(\xi)\cot^2(\phi)}.
\end{equation*}

Finally, the length of the arc $ \arc{BF} $ is given by 
\begin{align*}
	\nu(\arc{BF}) &=2\rho\arcsin(\sqrt{1-\cos^2(\xi)\csc^2(\phi)}).
\end{align*}
where $\cos(\xi) = {(\rho^2+r_e^2-d^2)}/{(2\rho r_e)}. $
\end{IEEEproof}
\begin{figure*}
	 	 \begin{align}
		\bP(D>d)&= \exp\left(-{\lambda}\int_{r_a}^{r_b}\int_{0}^{\xi}\left(1-e^{-\mu\pi^{-1}\arcsin\left(\sqrt{1-\cos^2(\xi)\sec^2(\varphi)}\right)}\right)\cos(\varphi)\diff \varphi \nu(\diff\rho)  \right)\label{eq:12},\\
					\bP(D=\infty)&=\exp\left(-{\lambda}\int_{r_a}^{r_b}\int_{0}^{\arccos(r_e/\rho)}\left(1-e^{-\mu\pi^{-1}\arcsin\left(\sqrt{1-{{r_e^2\sec^2(\varphi)/\rho^2}}}\right)}\right) \cos(\varphi)\diff \varphi \nu(\diff\rho) \right),\label{eq:15}\\
				\cL(f) & =\exp \left(-\frac{\lambda}{2\pi}\int_{r_a}^{r_b}  \int_0^\pi  \left(1-e^{-\frac{\mu}{2\pi}\int_{0}^{2\pi} 1-\exp{(-\bar{f}(\rho,\theta,\phi,\omega))}\diff \omega}\right) \sin(\phi) \diff\phi  \nu(\diff\rho) \right),\label{eq:16}\\
		\cL_{\Psi}(f)_{f=sH\|X-U\|^{-\alpha}}&= \exp \left(-\frac{\lambda}{2\pi}\int_{\bar{\cC}} \left(1-e^{-\frac{\mu}{2\pi}\int_{\bar{\omega}} 1-\cL_H(s(\rho^2-2\rho r_e\sin(\omega)\sin(\phi)+r_e^2)^{-\frac{\alpha}{2}})\diff \omega}\right) \sin(\phi) \diff\phi \diff\theta \nu(\diff\rho)  \right).\label{eq:totalint}
	\end{align}
\rule{\linewidth}{.1mm}
\end{figure*}
In downlink LEO satellite communications, network users are likely to get messages from their closest satellites. Let $D$ be the distance from the typical observer to its closest visible satellite, with $ D\stackrel{def}{=}\infty $ if there is no visible satellite.
 \begin{lemma}\label{Lemma:5}
The CDF of $ D $ is given by Eq. \eqref{eq:12} where $\cos(\xi)=(\rho^2+r_e^2-d^2)/(2\rho r_e). $

 \end{lemma}
\begin{IEEEproof}
	For $ r_a-r_e\leq d \leq \sqrt{r_b^2-r_e^2}, $ we have 
	\begin{align}
		\bP(D>d) &\ea \bP( \|X-u\|>d, \ \forall X\in\Psi)\nnb\\
		&\eb\bP(\|X_j-u\|>d,\ \forall  X_j\in\psi_{i},  \forall Z_i\in\Xi  )\nnb\\
		&=\bP\left(\prod_{Z_i\in\Xi}\bP\left(\left.\prod_{X_j\in\psi_{i}} \|X_j-u\|>d\right|\Xi\right)\right).\nnb
	\end{align}
	To get (a), we use the fact that for $ R>r, $ all satellites should be at distances greater than $ r. $ We have (b) by using that the Cox satellite point process is comprised of the Poisson point processes conditionally on orbits. We have 
	\begin{align}
		&\bP\left(\left.\prod_{X_j\in\psi_{i}} \|X_j-u\|>r\right|\Xi\right)\nnb\\
		&=\exp\left(-\mu\pi^{-1}\arcsin\left(\sqrt{1-{\cos^2(\xi)}{\csc^2(\phi_{i})}}\right)\right),\nnb
	\end{align}
where 
$ 	\cos(\xi)=(\rho_{i}^2+r_e^2-d^2)/(2\rho_{i}r_e)$.
We use the facts that (i) in order to have no point at distance less than $ r, $ the arc created by the orbit $ l(\rho_{i},\phi_{i},\theta_{i}) $ and the set $ C(\rho_i,d) $ has to be empty of satellite points and (ii) the void probability of the Poisson point process of intensity $ \mu $ on the arc is given by the $\exp(-\mu \cdot \text{arc length})$. Leveraging the fact that only the orbits with inclinations $ \pi/2-\xi<\phi<  \pi/2+\xi $ meet the spherical cap $ C(d) $, we have 
	\begin{align}
		&\bP(D>d)\nnb\\
		&=\bP\left(\prod_{{Z_i\in\Xi}}^{\pi/2-\xi<\phi< \pi/2+\xi}		e^{-\mu\pi^{-1}\sin^{-1}\left(\sqrt{1-\cos^2(\xi)\csc^2(\phi_{i})}\right)}\right)\nnb\\
		&= e^{\left.-{\lambda}\int_{r_a}^{r_b}\int_{0}^{\xi}\right.\left.(1-e^{-\frac{\mu}{\pi}\arcsin\left(\sqrt{1-{\cos^2(\xi)}{\sec^2(\varphi)}}\right)})\cos(\varphi)\diff \varphi \nu(\diff \rho) \right.},\nnb
	\end{align}
where we use the probability generating functional (pgfl) of $ \Xi $. Then, we employ the change of variable $\phi = \pi/2-\varphi$ with $ \cos(\xi)=(\rho^2+r_e^2-d^2)/(2\rho r_e). $ 
\end{IEEEproof}
\begin{definition}
	Outage occurs if the typical user has no visible satellite, or equivalently,
if $ D=\infty. $ \end{definition}
\begin{lemma}\label{lemma:outage}
	The outage probability is given by Eq. \eqref{eq:15}.
\end{lemma}
\begin{IEEEproof}
	There is no visible satellite iff $ D=\infty. $ By using Lemma \ref{Lemma:5}, the outage probability is given by 
	\begin{align*}
		&\bP(D=\infty)\\
		&= \bP(\|X_j-u\|>\sqrt{\rho_i^2-r_e^2}, \forall X_j\in\psi_i,\forall Z_i\in\Xi)\\
		&=\bP\left(\prod_{Z_i\in\Xi}\bP\left(\left.\prod_{X_j\in\psi_{i}} \|X_j-u\|>\sqrt{\rho_{i}^2-r_e^2}\right|\Xi\right)\right),\nnb
	\end{align*}
where we employ the pgfl of the Poisson point process of intensity $\mu$ to obtain the following expression:
	\begin{align}
	&\bP\left(\left.\prod_{X_j\in\psi_{i}} \|X_j-u\|>\sqrt{\rho_{i}^2-r_e^2}\right|\Xi\right)\nnb\\
	&=\exp\left(-\mu\pi^{-1}\arcsin\left(\sqrt{1-{r_e^2\csc^2(\phi_{i})}/{\rho_{i}^2 }}\right)\right).\nnb
\end{align}
We use the fact that when $ d=\sqrt{\rho_{i}^2-r_e^2} ,$ $ \cos(\xi)= r_e/\rho_{i}. $ 

For a given $ \rho, $ the orbits with inclinations $\pi/2-\xi<\phi< \pi/2+\xi$ meet the spherical cap $ C(\rho,\sqrt{\rho^2-r_e^2}) $.  Then, the outage probability is given by 
	\begin{align*}
	&\bP\left(\prod_{Z_i\in\Xi}e^{-\mu\pi^{-1}\arcsin(\sqrt{1-r_e^2\csc^2(\phi_i)/\rho_i^2})}\right)\\
	&=\exp\left(-{\lambda}\int_{r_a}^{r_b}\int_{0}^{\arccos(r_e/\rho)}\right.\\
	&\hspace{8mm}\left.\left(1-e^{-\frac{\mu}{\pi}\sin^{-1}\left(\sqrt{1-{{r_e^2\sec^2(\varphi)/\rho^2}}}\right)} \right)\cos(\varphi)\diff \varphi \nu(\diff \rho) \right),
\end{align*}
where we use a change of variables and the pgfl of $ \Xi $. 
\end{IEEEproof}
Fig. \ref{fig:outagevariable70007500} shows the outage probability obtained by Lemma \ref{lemma:outage}. 

	\begin{lemma}\label{Lemma:7}
		Consider a function $ f(X):\bR^3\to\bR^+ $. The Laplace functional of the proposed Cox point process $\cL_{\Psi}(f) = \bE_{\Psi}\left[\exp\left(-\sum_{X_i\in\Psi} f(X_i)\right)\right]$ is given by Eq. \eqref{eq:16}.
	\end{lemma} 


	\begin{IEEEproof}
		The Laplace functional of the proposed Cox point process is given by 
		\begin{align*}
			&\cL_{\Psi}(f)\\
			&= \bE\left[e^{-\sum_{X\in\Psi} f(X)}\right]\\
			&=\bE_{\Xi}\left[\bE_{\psi}\left[\left.e^{-\sum_{Z_i\in\Xi}\sum_{X_j\in\psi_{i}} f(X)}\right|\Xi\right]  \right]\\
			&=\bE_{\Xi}\left[\prod_{Z_i\in\Xi}\!\!\exp\left(-\frac{\mu}{2\pi}\int_{0}^{2\pi}1-e^{-\bar{f}(\rho_{i},\theta_{i},\phi_{i},\omega)}\diff \omega \right)  \right]\\
			&=\exp \left(-\frac{\lambda}{2\pi}\int_{r_a}^{r_b}\int_{0}^{\pi}\int^{\pi}_{0}\right.\\
			&\hspace{3mm}\left. \left(1-e^{-\frac{\mu}{2\pi}\int_{0}^{2\pi} 1-\exp{(-\bar{f}(\rho,\theta,\phi,\omega))}\diff \omega}\right)\sin(\phi)\diff \phi \diff \theta \nu(\diff \rho)\right), 
		\end{align*}
	where we use the function $ \bar{f}(\rho,\theta,\phi,\omega) {=} f(X)$ for any satellite $ X $ on the orbit $ l(\rho,\theta,\phi) $ with orbital angle $ \omega. $ Then, we use the pgfl  of the Poisson point process $\Xi $.  
	\end{IEEEproof}
	Consider a random variable $ H $ modeling random fading. Provided that the received signal power is $1$ at the reference distance, the received signal power at $ u  $ is $  f(X)=H\|X-u\|^{-\alpha}, $ where $ X $ is the location of the satellite and $ \alpha $ is the path loss exponent. The total interference power $ S $ is then given by the sum of the received signal powers from all the visible satellites as follows: 
\begin{align}
	S &= \sum_{X\in\bar{\Psi}} H\|X-u\|^{-\alpha},
		\bar{\Psi}= \Psi\left(\bigcup\limits_{r_a<\rho\leq r_b} \!\!\!\! C(\rho,\sqrt{\rho^2-r_e^2})\right). \nnb
\end{align} 
 
\begin{corollary}
	The Laplace transform of the total interference is given by Eq. \eqref{eq:totalint} where $ \bar{\cC}=\{(\rho,\theta,\phi)\in \cC \text{ s.t. } l(\rho,\theta,\phi)\cap C(\sqrt{r_b^2-r_e^2})\neq \emptyset \} $ and $\bar{\omega} = \{\omega\in [0,2\pi]| X(\rho,\theta,\phi,\omega)\in C(\sqrt{r_b^2-r_e^2}), \forall (\rho,\theta,\phi)\in\bar{\cC} \} $.
\end{corollary}
\begin{IEEEproof}
	The Laplace transform is given by 
	\begin{align*}
		&\cL_{\Psi}(f)_{f=sH\|X-U\|^{-\alpha}}\\
&=\bE_{\Xi}\left[\prod_{{Z_i\in\Xi}}\bE_{\psi}\left[\left.\prod_{X_j\in\psi_{i}}\cL_{H}(s\|X_j-u\|^{-\alpha}) \right.\right]  \right],
	\end{align*}
where $ \cL_{H}(\kappa) $ is the Laplace transform of the fading random variable $ H$. Using a technique similar to Lemmas \ref{lemma:3} and \ref{Lemma:7}, we obtain the final result. 
	\end{IEEEproof}	

	\section{Numerical Results}
	\subsection{Starlink and Cox Model}
The proposed analytical framework offers a mathematical representation of both existing and future LEO satellite networks. Here, we illustrate the applicability of the proposed model by using it to model the prospective Starlink 2A network \cite{FCC}, which consists of 28 orbital planes and 120 satellites per plane. The altitudes of the orbital planes are 525 km, 530 km, and 535 km, and their inclinations are 53 $\degree$, 43$\degree$, and 33$\degree$, respectively. 

To represent this Starlink network based on our Cox model, we adopt a moment matching method where we adjust the local density parameters $\lambda$, $\mu$, to ensure that the average numbers of LEO satellites are the same for the Starlink and the Cox constellations. For satellite altitude, we use $(r_a, r_b) = (6896 \text{km}, 6906 \text{km})$. We then use the same assumptions for both cases, such as nearest association, Nakagami-m fading with $\alpha=2$, a frequency spatial reuse factor of 4\footnote{In the Starlink, one out of every four satellites on each orbit uses the same spectrum; In the Cox, the Poisson point process of intensity $\mu/4$ satellites on each orbit use the same spectrum.}, a carrier frequency of 6 GHz, a bandwidth of 10 MHz, and an antenna gain of 20 dB from the association satellite. 

Figure \ref{fig:comparisoncppbpp} shows the simulated SIR coverage probability of the typical user in the Starlink network and the SIR coverage probability of the typical user in the moment-matched Cox network. The figure demonstrates that our proposed Cox model closely approximates the coverage probability of the forthcoming Starlink constellation. This highlights the potential of our analytical model in effectively approximating the network performance of future LEO satellite constellations.

\subsection{Cox and Binomial Point Processes}

\begin{figure}
		\centering
		\includegraphics[width=1\linewidth]{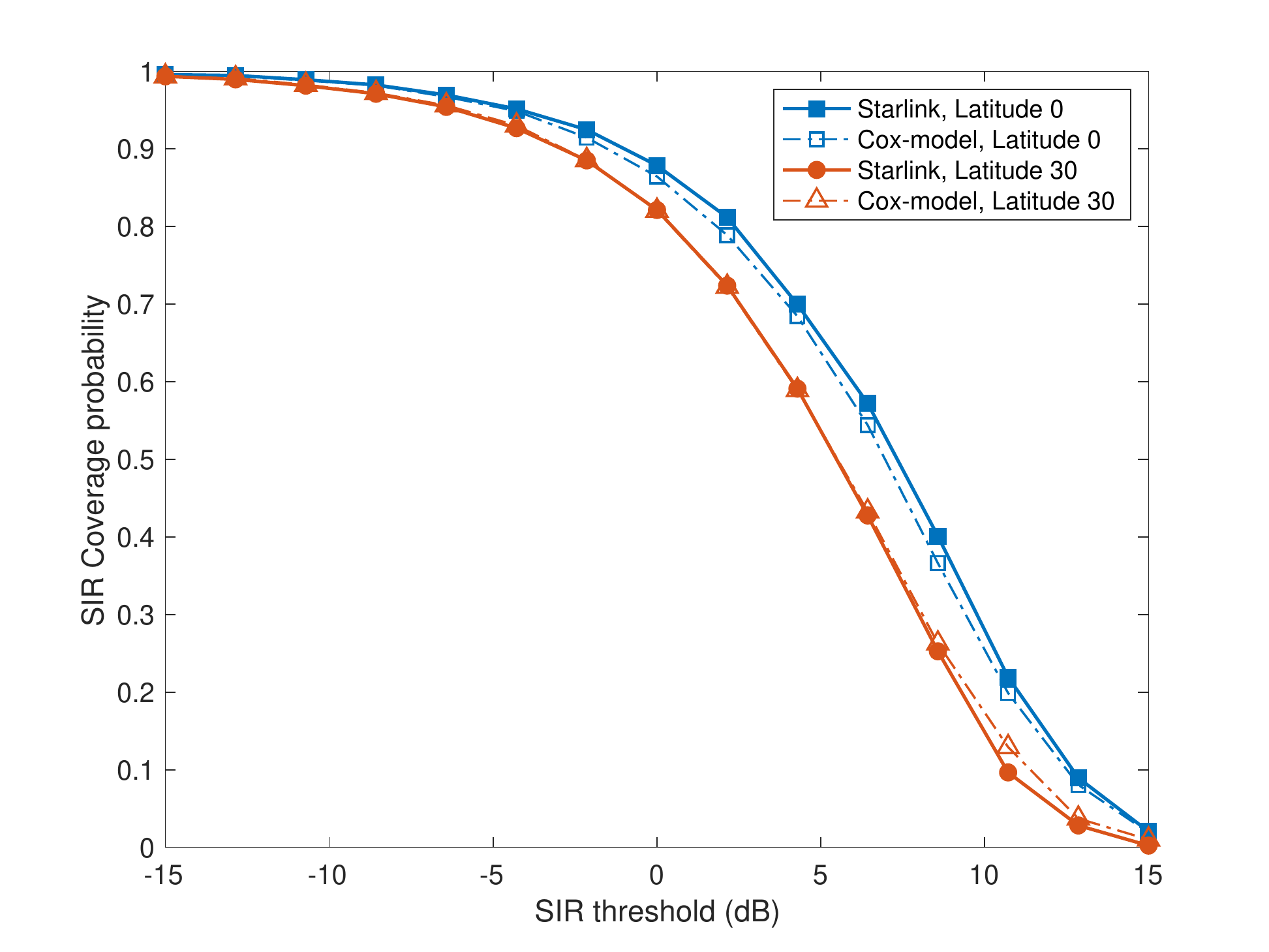}
		\caption{The SIR coverage probability of the typical user at latitudes $0$ and $30\degree$. For simplicity, we use Nakagami-$m$ fading. The Cox model at latitude $0$ means that we use the best fit isotropic Cox point process that approximate the SIR of the Starlink.}
		\label{fig:comparisoncppbpp}
	\end{figure}

\begin{figure}
		\centering
		\includegraphics[width=1\linewidth]{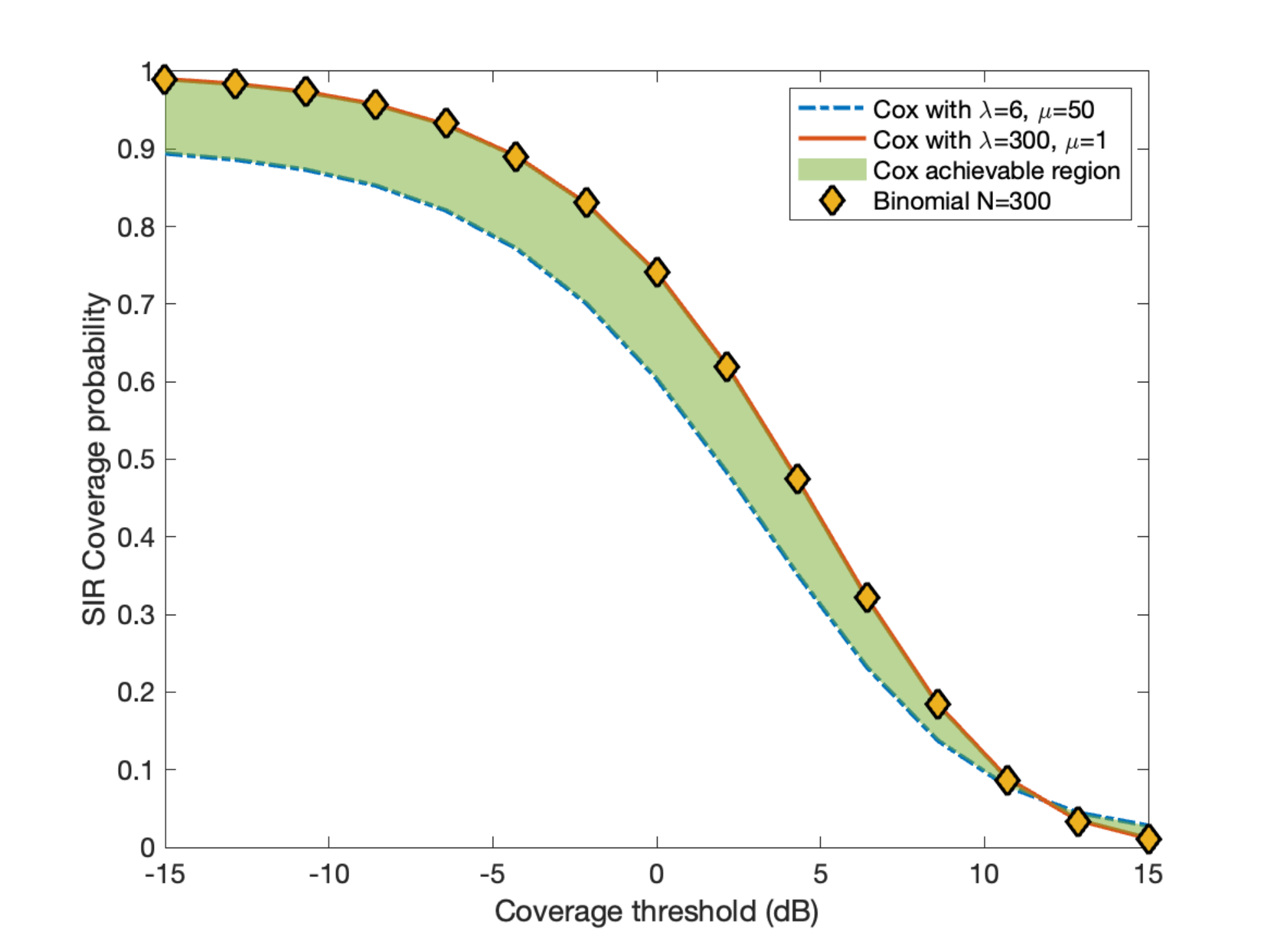}
		\caption{SIR coverage probability for the Binomial and Cox models. We assume $g=10$ dB, the antenna gain from the nearest satellite.}
		\label{fig:binomialvscoxg10}
	\end{figure}

The developed Cox model is flexible enough to produce diverse network layouts. Specifically, for any LEO satellite constellation, one can only adjust the parameter \( N \) of the binomial point process to represent it. In contrast, for the same LEO constellation, one can jointly adjust \( \lambda \) and \( \mu \) to describe it. There are numerous ways to generate a LEO satellite constellation by varying the numbers of orbits and the mean number of satellites per orbit while keeping $N=\lambda\mu$. In short, to depict an existing or future constellation, the proposed Cox model produces a larger number of network layouts, leading to a much more comprehensive network performance analysis compared to the analysis based on the binomial point process. This property of the Cox model is well depicted in Fig. \ref{fig:binomialvscoxg10} where various SIR coverage probabilities are numerically evaluated for (i) Cox with \( \lambda = 6, \mu = 50 \), (ii) Cox with \( \lambda = 300, \mu = 1 \), (iii) Cox by varying \( \lambda \) and \( \mu \) such that \( \lambda \mu = 300 \), and (iv) binomial with \( N = 300 \). To evaluate the SIR coverage, we reuse the assumption used for Fig. \ref{fig:comparisoncppbpp}. It is worthwhile to note that in all cases, the total number of satellites is the same and equal to $300$. The figure clearly demonstrates that by allowing numerous network layouts through adjusting \( \lambda \) and \( \mu \), the developed Cox model produces a wide range of SIR coverage probabilities, illustrated by the green area in Fig. \ref{fig:comparisoncppbpp} which represents the Cox achievable region.

	\section{Conclusion}
This paper builds a novel stochastic geometry framework for modeling the spatial distribution of clustered LEO satellites along orbits of varying altitudes, obtained through the development of an isotropic satellite Cox point process. The paper explicitly provides expressions for essential statistical properties, including the distribution of nearest distances to LEO satellites, outage probability, Laplace function of the Cox point process, and Laplace transform of the total interference experienced by a typical user. These outcomes can be directly employed to assess the typical performance of future multi-altitude LEO satellite downlink communications.



\section*{Acknowledgment}
 This paper is a joint study with the support of Korea-France Cooperative Development Program (STAR) from the NRF, funded by the Korean government in the year 2024 (RS-2024-00247682). The work of Chang-Sik Choi was supported by the NRF-RS-2024-0034240. The work of Francois Baccelli was supported by ERC NEMO grant 788851 to INRIA and by the French National Agency for Research (ANR) via the project n°ANR-22-PEFT-0010 of the France 2030 program PEPR.
 {\em Réseaux du Futur}.
	\bibliographystyle{IEEEtran}
	\bibliography{ref}

\end{document}